# Dynamic actuation of DNA-assembled plasmonic nanostructures in microfluidic cell-sized compartments

Kerstin Göpfrich,[*,†,‡,¶] Maximilian J. Urban,[§,¶] Christoph Frey,[⊥,#] Ilia Platzman,[⊥,#] Joachim P. Spatz,[*,⊥,@,#] and Na Liu[*,§]

†Max Planck Institute for Medical Research, Biophysical Engineering Group, Jahnstraße 29, 69120 Heidelberg, Germany

‡Department of Physics and Astronomy, Heidelberg University, 69120 Heidelberg, Germany

¶Equal contribution.

§Max Planck Institute for Intelligent Systems, Heisenbergstraße 3, 70569 Stuttgart, Germany

Kirchhoff Institute for Physics, Heidelberg University, Im Neuenheimer Feld 227, 69120 Heidelberg, Germany

⊥Max Planck Institute for Medical Research, Department of Cellular Biophysics, Jahnstraße 29, 69120 Heidelberg, Germany

#Department of Biophysical Chemistry, Heidelberg University, Im Neuenheimer Feld 253, 69120 Heidelberg, Germany

@Max Planck School Matter to Life, Jahnstraße 29, 69120 Heidelberg, Germany

E-mail: kerstin.goepfrich@mr.mpg.de; spatz@mr.mpg.de; na.liu@kip.uni-heidelberg.de

# ABSTRACT

Molecular motor proteins form the basis of cellular dynamics. Recently, notable efforts have led to the creation of their DNA-based mimics, which can carry out complex nanoscale motion. However, such functional analogues have not yet been integrated or operated inside synthetic cells towards the goal of realizing artificial biological systems entirely from the bottom-up. In this Letter, we encapsulate and actuate DNA-assembled dynamic nanostructures inside cell-sized microfluidic compartments. These encapsulated DNA nanostructures not only exhibit structural reconfigurability owing to their pH-sensitive molecular switches upon external stimuli, but also possess optical feedback enabled by the integrated plasmonic probes. In particular, we demonstrate the power of microfluidic compartmentalization for achieving on-chip plasmonic enantiomer separation and substrate filtration. Our work exemplifies that the two unique tools, droplet-based microfluidics and DNA technology, offering high precision on the microscale and nanoscale, respectively, can be brought together to greatly enrich the complexity and diversity of functional synthetic systems.

## Keywords

The futuristic objective of synthetic biology is to build artificial systems such as synthetic cells, in which different functional entities can be designed and constructed entirely from the bottom-up.[1–3] Progress towards this objective requires inspiring strategies to combine and arrange a multitude of functional building blocks in space and time. To this end, microfluidic technologies offer unique on-chip modules for synthetic cell assembly and subsequent manipulation by deformation,[4] splitting,[5,6] pico-injection,[7,8] fusion[9] or content mixing.[10] Automated analysis in trapping chambers,[11] sorting[9] and content-release[8] modules further push the potential of existing experimental schemes and opportunities. In particular, the combination of these modules has enabled the development of semi-automated sequential assembly lines for synthetic cells.[8] However, the operation of complex subcellular functions and function-chains inside cell-sized compartments generally relies on natural cellular components, such as transmembrane or cytoskeletal proteins, among others. Their reconstitution in cell-sized compartments has led to the realization of intricate subcellular features, such as minimal actin cortex assembly[12–14] or microtubule networks[15] in synthetic cells. However, protein purification can be challenging and one may argue that a truly synthetic cell should contain solely man-made components.

In recent years, DNA nanotechnology, particularly DNA origami, has embarked on the design of protein mimics – from DNA-based ion channels[16–18] to scramblases,[19] membrane-bending structures[20,21] and, remarkably, artificial motors.[22,23] It has also been possible to construct reconfigurable DNA origami structures with tailored optical functionalities.[24–26] However, so far actuation and operation of dynamic DNA origami structures have not yet been demonstrated inside cell-like compartments for advancing the field further forward. In this Letter, we realize this key objective by encapsulating DNA-assembled reconfigurable nanostructures inside microfluidic compartments. The reversible operation of such nanostructures inside the compartments is triggered by external stimuli and can be directly read out by optical spectroscopy. We further showcase the potential of compartmentalization in combination with the programmability of DNA-based nanosystems to achieve complex and highly spe-cific separation functions for plasmonic enantiomer selection in microfluidic compartments.

The reconfigurability of the DNA-assembled nanostructures can be enabled by functionalization with various molecular switches, which respond to specific external stimuli, such as proteins, aptamers, light, temperature or ions. In our case, a DNA origami cross structure, which can reconfigure upon pH changes, is designed and implemented as shown in Figure 1a. It comprises two stiff DNA origami arms linked by a flexible hinge at its centre. The two arms can be locked or unlocked by a pH-sensitive DNA switch based on a triplex motif.[26,27] Each arm of the cross is functionalized with a gold nanorod (AuNR), allowing for optical detection of the conformational changes inside a microfluidic droplet by circular dicroism (CD) spectroscopy. In addition, the AuNRs help to readily evaluate the structural integrity with transmission electron microscopy (TEM) prior to encapsulation as well as after release from the microfluidic droplets. The assembled DNA origami nanostructures are injected into a microfluidic PDMS-based droplet formation device via the aqueous inlet as illustrated in Figure 1b. Cell-sized water-in-oil compartments are formed at the T-junction of the microfluidic device, where the aqueous phase (containing ∼ 1 nM DNA origami nanostructures in 1×TAE, 14 mM $MgCl_2$) is intersected by an oil phase (HFE-7500 supplemented with 2 wt% of a PEG-based fluorosurfactant, see Methods; for the detailed layout of the microfluidic device, see Supporting Information, Figure S1). The encapsulation of the nanostructures leads to the formation of plasmonic droplets. The surfactant serves as a droplet-stabilizing agent, preventing the fusion of the droplets when densely packed.[8] The droplet formation process is monitored by an inverted microscope equipped with a high-speed camera. A representative image of the process is shown in Figure 1c.
After the droplet formation, confocal fluorescence microscopy is employed to verify whether the DNA origami structures have been successfully encapsulated and stably remain within the droplet lumen for subsequent experiments. Figure 1d shows a representative brightfield image of the droplets, confirming their uniform size (radius: 25.8 ± 1.5 $\mu$m, n = 54).

The corresponding fluorescence image (see Figure 1e) validates the confinement of the DNA nanostructures (stained with the intercalating dye SYBR Green I) inside the droplets. They are homogeneously distributed within the droplet lumen and no aggregates observable in Figure 1e. The image has been taken 24 h after the droplet formation, demonstrating good stability of the droplets as well as successful confinement of the DNA nanostructures inside the droplets. The fluorescence intensity inside the droplets is directly proportional to the number of encapsulated plasmonic nanostructures. A mean fluorescence intensity of 89.8 a.u. ± 5.3 a.u. (mean ± std., $n$ = 58) has been achieved, demonstrating homogeneous encapsulation (see Supporting Information, Figure S2).

pH gradients are the universal energy source inside living cells. Alterations of the intra- or extracellular pH values are often markers of disease.[28] To implement dynamic pH responses, the DNA switch that links the two arms of the DNA origami cross is modified with a triplex motif.[26] At pH ≤ 8.5, the single-stranded DNA on one arm binds to the DNA duplex positioned on the other arm via Hoogsten interactions. A DNA triplex forms. This gives rise to the locked state of the DNA origami cross as illustrated in Figure 2a. At pH > 9, dissociation of the DNA triplex results in the unlocked state of the cross. While in bulk, such conformational changes can be triggered simply by adding $H^+$ or $OH^-$, respectively, this strategy is no longer straightforward after compartmentalization. We hence have developed a non-invasive approach to increase the pH inside the droplets without interfering with the droplet integrity by addition of pyridine to the oil phase of the droplet emulsion as illustrated in Figure 2b. Once pyridine reaches the droplet interface, it acts as a proton acceptor. Protons are depleted from the aqueous phase of the droplet, leading to a pH increase inside the droplet and hence to the unlocking of the DNA origami crosses. Subsequent addition of a proton donor (e.g. Krytox) drives the origami structures back to the initial locked state. To test this strategy, CD characterization of the plasmonic droplets has been carried out. Due to the high optical density of the surrounding oil phase, a quartz cuvette with a short optical path length (0.1 mm) is used to obtain optical signals with a good signal-to-noise ratio. As shown in Figure 2c, a characteristic dip-to-peak spectral profile is observed at pH = 8.5, indicating a right-handed (RH) locked state of the plasmonic crosses.[29] Addition of pyridine leads to an apparent CD signal decrease, resulting from the opening of the plasmonic crosses. Subsequent addition of Krytox recovers the CD signals corresponding to the locked state. As shown in Figure 2c, the dynamic actuation can be operated in multiple cycles. It is noteworthy that the AuNRs are utilized to achieve a chiroptical response, ideally suited to optically monitor the pH changes inside the droplets. In principle, the conformational switch also takes place in the absence of the AuNRs, as it is triggered by the DNA triplex motif alone.[26,27] To demonstrate the structural integrity of the DNA-assembled plasmonic crosses after the droplet encapsulation and pH cycling, the structures are released from the droplets and subjected to TEM characterization. Figure 2d shows the TEM image of the structures after a full pH cycle. It is evident that the two AuNRs are stably assembled on the individual origami crosses.

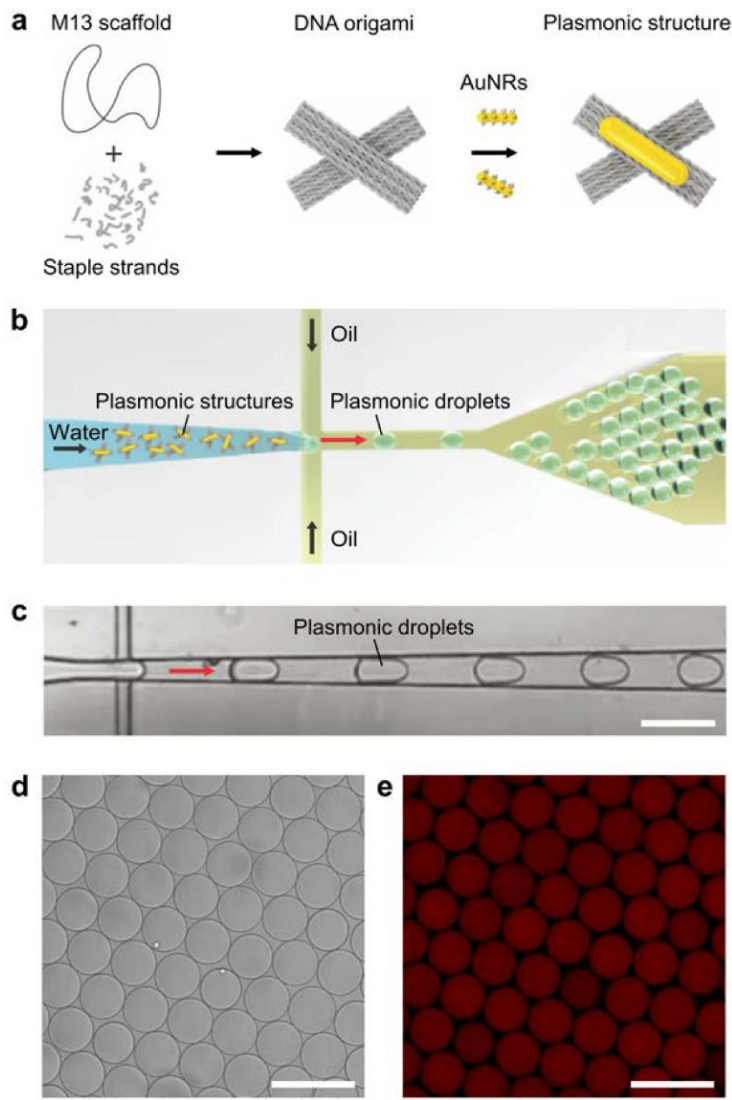

Figure 1: Encapsulation of the DNA-assembled plasmonic nanostructures in microfluidic droplets. a) Self-assembly of the DNA-assembled plasmonic nanostructure. A DNA origami cross is assembled from the p7560 scaffold and staple strands. It is then functionalized with one AuNR on each arm of the cross to form a plasmonic nanostructure. b) Schematic of the microfluidic encapsulation of DNA-assembled plasmonic nanostructures in water-in-oil droplets. Plasmonic droplets are formed at the microfluidic T-junction c) High-speed microscopy image (brightfield) of the encapsulation process. DNA-assembled plasmonic nanostructures are encapsulated via the aqueous channel of a microfluidic droplet formation device. Scale bar: 100 $\mu$m. d) Brightfield and e) Confocal fluorescence image of the plasmonic droplets 24 h after encapsulation (stained with SYBR Green I, $\lambda_{ex}$ =488 nm). Scale bar: 100 $\mu$m.

Next, we set out to harness the powerful combination of DNA nanotechnology and droplet-based microfluidics to achieve novel functions. In particular, we demonstrate an on-chip filtration function to implement plasmonic enantiomer selection and separation. To this end, we encapsulate cholesterol-tagged DNA together with a racemic mixture of right handed (RH) and left handed (LH) plasmonic crosses. The cholesterol-tagged DNA self-assembles into a surfactant layer at the compartment periphery due to hydrophobic interactions.[12] It serves as a programmable anchor to exclusively recruit the LH species functionalized with a complementary DNA overhang to the compartment periphery as highlighted in Figure 3a. The RH structures, on the other hand, remain homogeneously distributed within the droplets in the aqueous phase. Plasmonic enantiomer separation is achieved by breaking up the droplet emulsion: The RH structures are released, whereas the LH structures remain bound to the water-oil interface as illustrated in Figure 3a. Note that for efficient filtra-tion, it is necessary to use an excess of the cholesterol-tagged DNA to present sufficient attachment points for the LH species (see Supplementary Information, Figure S4 and Text S1). For controlled destabilization of the emulsion and subsequent release, a microfluidic chip architecture is employed as illustrated in Figure 3b. At a microfluidic y-junction, the droplet-filled oil channel encounters an aqueous phase (0.5×TAE, 11 mM $MgCl_2$). The co-flowing immiscible fluids pass an inbuilt electrode for the application of an external electric field (500 V AC). Once the electric field is turned on, dielectrophoretic forces cause the fusion of the droplets at the water-oil interface, releasing their content into the continuous aqueous phase. Figure 3c shows a brightfield image of the release section of the device. Using a high speed camera, multiple droplets are captured while they approach the electrodes and ultimately release their content into the aqueous phase. The content of up to 200 droplets can be released per second. When the electric field is turned on, release efficiencies of 100 % are achieved, meaning all the droplets fuse with the aqueous phase. This demonstrates the fidelity of our chip design and overall the advantages of using microfluidics as a high-precision tool. Figure 3d shows a confocal fluorescent image of the microfluidic droplets. The LH structures (cyan) colocalize with the droplet periphery due to complementary base-pairing with the cholesterol-tagged DNA. It is apparent that the RH structures stay in the droplet lumen (red). The crossectional fluorescence intensity profiles confirm the different distributions of the two species. To demonstrate the feasibility and effectiveness of the plasmonic enantiomer selection process, CD measurements are carried out before encapsulation and after the droplets pass through the microfluidic release device. The racemic mixture exhibits low CD signals, because the contributions from the LH and RH structures nearly cancel out, as presented by the grey curve in Figure 3f. After the selection process, the CD result characterized by the red curve clearly confirms the recovery of the RH signal. The distinct CD spectrum also indicates good structural integrity of the plasmonic crosses after passing through the electric field on the microfluidic device. It is noteworthy that a similar selection process could, in principle, be performed at any surfactant-stabilized water-oil in-terface. The advantage of compartmentalization lies in the increased surface area, which can greatly improve the filtration efficiency. Assuming an aqueous phase of 100 µl encapsulated into droplets with a radius of 20 µm, a notably large surface area of about 150 cm$^2$ can be obtained. Additionally, the confinement in the droplets limits the diffusion time compared to the bulk experiment.

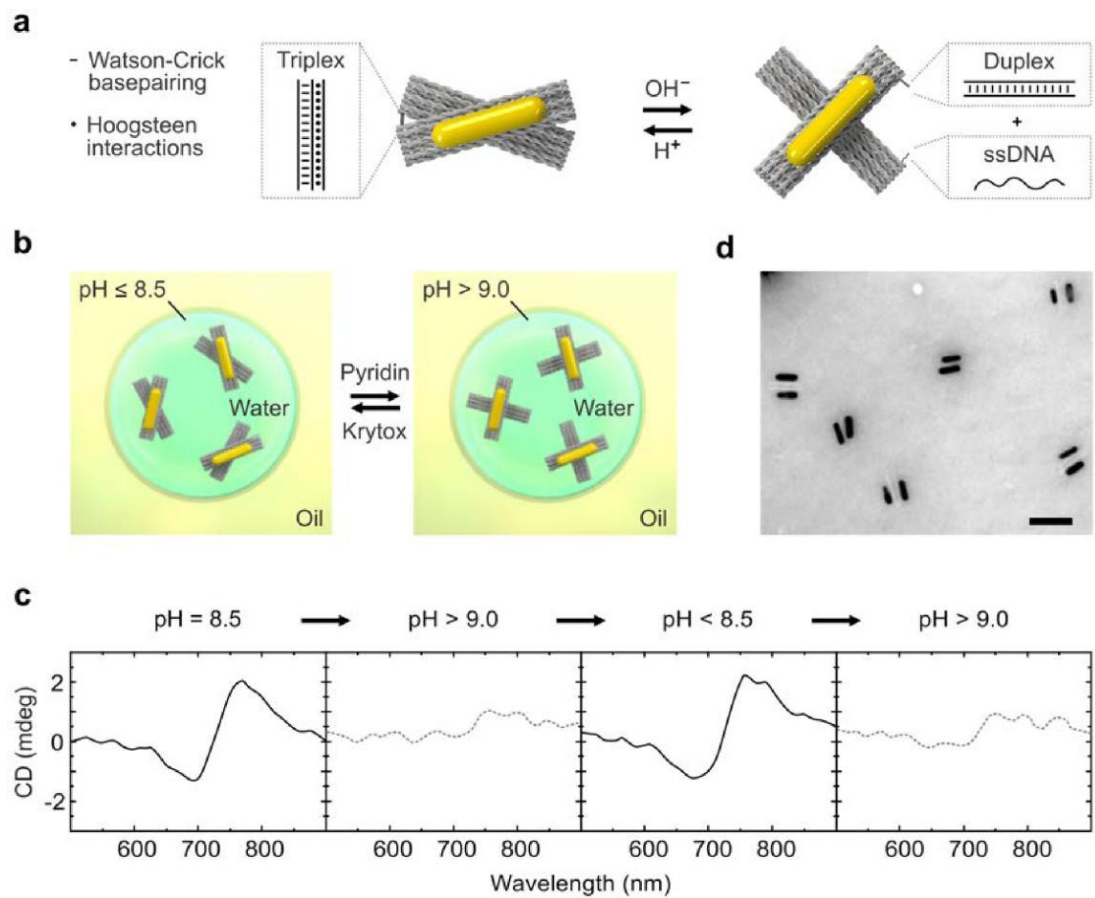

Figure 2: Reversible actuation of the DNA-assembled plasmonic nanostructures in cell-sized microfluidic compartments. a) Illustration of the the DNA-assembled plasmonic nanostructures modified with a pH-sensitive triplex motif. Hoogsten interactions lead to the locked state of the cross at pH ≤ 8.5. The tripex dissociates at higher pH values, leading to the unlocked state. b) Schematic of the non-invasive pH-responsive actuation of the droplet-encapsulated plasmonic crosses upon addition of the proton acceptor pyridine or the proton donator Krytox. c) CD spectra of the plasmonic droplets at different pH values, demonstrating good reversibility. d) TEM image of the DNA-assembled plasmonic nanostructures with two AuNRs each after encapsulation, pH cycling and subsequent release from the mircofluidic droplets. Scale bar: 100 nm.

By combining a microfluidic droplet release function with DNA nanotechnology, we hence achieve an efficient sorting and filtration process. Also, the described filtration mechanism is broadly applicable beyond the plasmonic enantiomer selection. In principle, any component can be sorted and isolated from the aqueous phase – from proteins to cells by coupling the respective molecular recognition site for the substrate of interest to the cholesterol-tagged DNA.

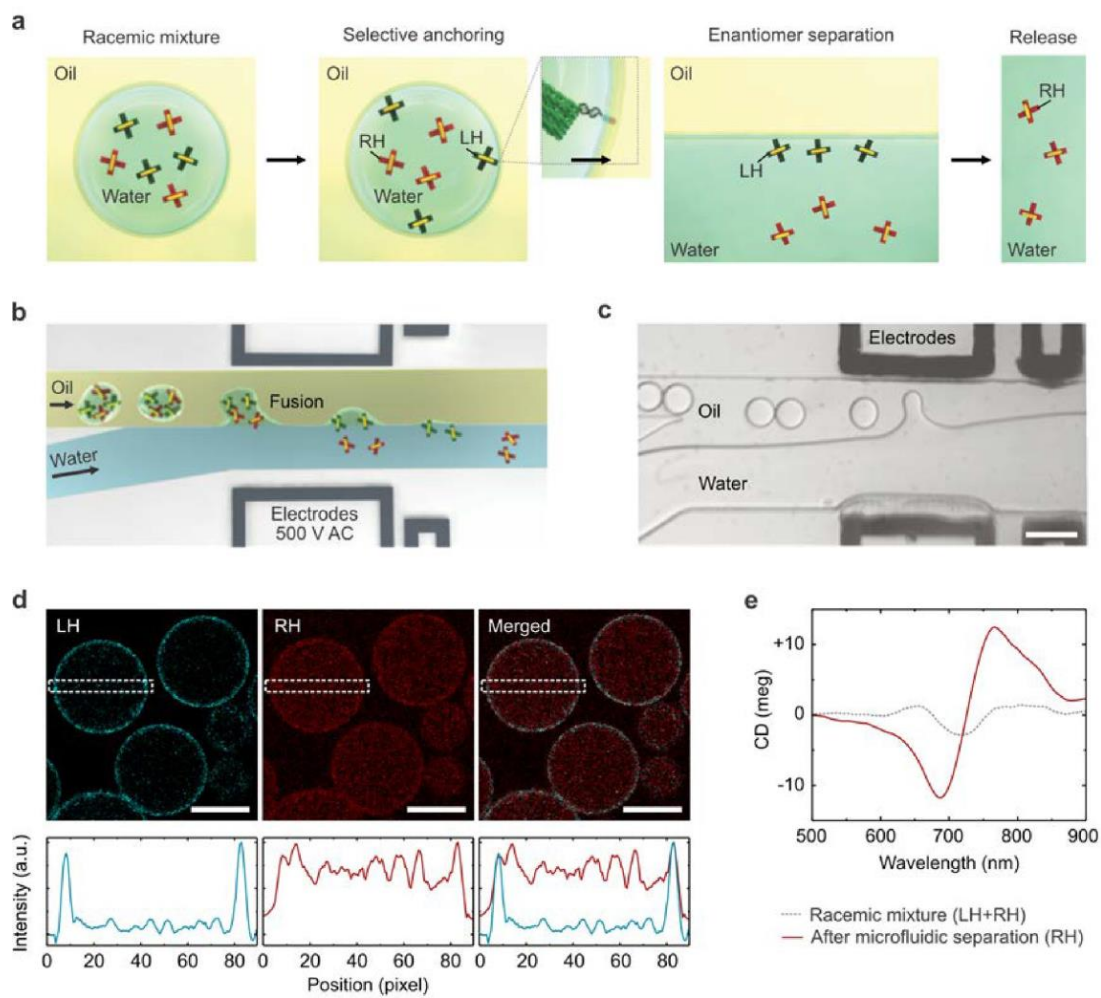

Figure 3: Microfluidic plasmonic enantiomer selection. a) Plasmonic enantiomer selection process. The left handed (LH) structures are selectively anchored to the droplet periphery via complementary cholesterol-tagged DNA. After release, they remain bound to the water-oil interface, while the right handed (RH) structures are in the bulk solution. b) Schematic of the microfluidic release device. Droplets are destabilized by an electrical field (500 V AC), releasing their content into the aqueous phase. c) High speed camera image (brightfield, scale bar: 100 $\mu$m) of the release process. d) Confocal fluorescence images of microfluidic droplets encapsulating the racemic mixture of the LH (Atto 550 labelled, $\lambda_{ex}$ =493 nm) and RH structures (Atto 647 labelled, $\lambda_{ex}$ =653 nm). Scale bar: 30 $\mu$m. The intensity profiles (bottom row) of the cross section regions (indicated by the white box) confirm the selective binding of the LH structures to the compartment periphery. e) CD spectrum before (grey curve) and after (red curve) the microfluidic selection process. The increase in the CD signal after the selection confirms the effective enrichment of the RH structures in solution.

One of the most exciting tasks towards the assembly of synthetic cells is the integration of molecular motors with the capacity to perform mechanical motion upon stimulation. In living cells, protein-based molecular machines accomplish literally any tasks, from energy generation to cell division, directed transport and cargo sorting. Their de novo reconstruction from DNA and their operation in synthetic cells is an extremely fruitful albeit challenging endeavor. Our study demonstrates how the combination of precision technologies – DNA nanotechnology and droplet-based microfluidics – provides new possibilities to construct, operate and monitor artificial dynamic nanostructures inside synthetic cells. We prove their reversible pH-triggered reconfiguration by CD spectroscopy, TEM and confocal fluorescence imaging. Promising potential for future work features the realization of stimulus cascades, mimicking reaction pathways in living cells. Furthermore, the integration of DNAzymes or catalytically active nucleic acids into droplets and liposomes will allow for increasingly complex functions. Such approaches will enable synthetic cellular systems, which are not merely copies of nature's own example of life, but prescribe a new direction, a new example of how life could be. Along the route, we may also discover fundamental design principles of living systems, which can be harnessed and extremely useful for biomedicine and technologies.

# Methods

## Design and fabrication of the microfluidic devices

Microfluidic PDMS (Polydimethylsiloxane, Sylgard 184, Dow Corining)-based devices for droplet production and content release were designed with QCAD-pro (RibbonSoft GmbH) and fabricated as previously described using photo- and soft-lithography.[8] For the droplet release device, electrodes were inserted into the pre-assembled microfluidic chip. For this purpose, the chip was placed on a hot plate at 80 °C and a low melting-point solder (RS Components Ltd.) was melted inside the empty electrode microchannels as previously described.[30] Electrical wires were connected to the melted solder. For the device layouts, see

Supporting Information, Figures S1 and S3.

## Design and assembly of the pH-responsive structures

The DNA origami structures were designed with caDNAno 2.0.[31] All DNA sequences are provided in the Supporting Information (Tables S2 – S12) and the routing of the staple strands is identical to a previously published structure.[26] For assembly, 10 nM p7560 scaffold (tilibit nanosystems GmbH) were mixed with 100 nM of each staple (purification: desalting, Eurofins Genomics), including those modified with sequences for pH locks (purification: high-performance liquid chromatography (HPLC), Sigma-Aldrich). The mixture was placed in a thermocycler (Eppendorf Mastercycler Pro, Merck KGaA) and annealed as described in the Supporting Information, Table S1. The LH and RH structures were assembled separately. After assembly, the excess staple strands were removed by gel electrophoresis (1.5 % agarose gel containing SYBR Safe DNA stain, 0.5× Tris Boric Acid EDTA buffer (TBE), and 11 mM $MgCl_2$). Target bands were cut out, and the DNA origami structures were extracted with Freeze'N Squeeze spin columns (Bio-Rad Laboratories GmbH).

## Attachment of the AuNRs

AuNRs (10 nm x 38 nm) were purchased from Sigma-Aldrich. They were first functionalized with thiolated DNA (SH-5' TTTTTTTTTTTTTTT 3', purification: HPLC, biomers.net GmbH) via the low-pH route as previously described.[32] The functionalized AuNRs were purified and assembled on the pre-folded DNA origami according to a published protocol.[24]

## Microfluidic encapsulation

For encapsulation of the DNA-assembled plasmonic structures in microfluidic compartments, a droplet production device (for detailed device layout see Supporting Information, Figure S1) was utilized. The device featured a T-junction where the fluid flow from the aqueous inlet is intercepted by an oil phase, leading to the formation of water-in-oil emulsion droplets. The droplet formation process was monitored with an inverted microscope (Axio Vert.A1, Carl Zeiss AG) equipped with a high-speed camera (Phantom v 2511, Vision Research) and a 20x objective (Carl Zeiss AG). The aqueous phase contained ~1 nM of the plasmonic nanostructures in 0.5× TBE and 11 mM $MgCl_2$. The oil phase consisted of 2 wt% Perflouro-polyether-polyethylene glycol (PFPE-PEG) block-copolymer fluorosurfactants (PEG-based fluorosurfactants, Ran Biotechnologies Inc.) dissolved in HFE-7500 oil (DuPont). At the outlet, droplets were collected in a microtube for further experiments.

## Confocal fluorescence microscopy

For the confocal imaging, a Zeiss LSM 800 confocal (Carl Zeiss AG) with a 20x air objective (Plan-Apochromat 20x/0.8 M27, Carl Zeiss AG) was used. The pinhole aperture was set to one Airy Unit and the experiments were performed at room temperature. The recorded images were brightness and contrast adjusted and analysed with ImageJ (NIH).

## TEM characterization

The DNA origami structures with the AuNRs were visualized using a Philips CM 200 TEM operating at 200 kV before and after encapsulation into microfluidic droplets. To release the structures from the droplets for imaging after encapsulation, the droplet emulsion was destabilized by adding 100 µL of perfluoro-1-octanol (PFO) destabilizing agent (Sigma-Aldrich). Within seconds to minutes, the milky emulsion broke up and disappeared, forming a transparent aqueous layer on top of the oil-surfactant mixture. This top layer contains the released DNA nanostructures, which were carefully removed with a pipette. For TEM imaging, the DNA structures were deposited on freshly glow-discharged carbon/formvar TEM grids. The TEM grids were treated with a uranyl formate solution (0.75 %) for negative staining of the DNA structures. Uranyl formate for negative TEM staining was purchased from Polysciences, Inc. Images were analysed in ImageJ (NIH).

## CD spectroscopy

CD spectroscopy was performed using a J1500 Circular Dichroism Spectrometer (JASCO). The droplets were pipetted into a Quartz SUPRASIL cuvette (Hellma GmbH & Co. KG) with an optical path length of 0.1 mm (see Figure 2). Note that due to the high optical density of the oil phase, it was crucial to choose a cuvette with a short optical path length to maximize the optical signals. For CD spectroscopy of the bulk solution (Figure 3), a cuvette with an optical path length of 1 cm was used. All measurements were carried out at room temperature.

## pH cycling experiments

To trigger pH-dependent conformational switching (see Figure 2b), 500 $\mu$L of the droplet emulsion was placed in a microtube (Eppendorf GmbH). 10 $\mu$l of the solution were removed for the initial CD measurement (see Figure 2b). Pyridine (Merck KGaA) was dissolved in HFE-7500 (DuPont) at a volumetric ratio of 1:1000. Pyridine (10 $\mu$l) was added to the droplet emulsion. Pyridine acted as a proton acceptor and hence reduced the pH inside the droplets.[8] After 10 min equilibration, another 10 $\mu$l of the droplet emulsion were removed for the second CD measurement (see Figure 2b). Subsequently, 10 $\mu$l of 100 mM PFPE-carboxylic acid (Krytox, MW: 7000-7500 g/mol, DuPont) in HFE-7500 were added to the remaining droplet emulsion. Krytox acted as a proton donor and hence caused a pH decrease inside the droplets.[33] This procedure was repeated to complete the pH cycling experiment, adjusting the Krytox / pyridine concentration in each cycle, presented in Figure 2b.

## Microfluidic plasmonic enantiomer selection

The LH structure was modified with seven ssDNA overhangs (see Supporting Information, Table S11), complementary to the sequence of a cholesterol-tagged DNA handle (5'TGAT-GCATAGAAGGAA/3CholTEG/3', purification: HPLC, Integrated DNA Technologies).

Droplets containing an enantiomeric mixture of the LH and the RH structures (1:1 ratio) as well as 1 µM of the cholesterol-tagged DNA were prepared with the droplet formation device as described. Plasmonic enantiomer selection was performed using the microfluidic droplet release device (for detailed device layout see Supporting Information, Figure S2). Droplets were injected via the oil inlet of the device, the aqueous phase consisted of buffer only (0.5x TBE, 11 mM $MgCl_2$). To trigger the release of the droplet content, an AC field of (500 V, 0.5 kHz) was applied via the electrode facing the aqueous side (note that the second electrode near the oil flow was inactive). After release, the aqueous phase was collected in a microtube for the subsequent experiments.

# ASSOCIATED CONTENT

**Supporting Information.**

The following files are available free of charge.

Layout of the microfluidic devices. Homogeneity of the encapsulation. Attachment efficiency and density of the cholesterol-tagged DNA. Annealing protocol for the DNA nanostructures. All DNA sequences (PDF)

# AUTHOR INFORMATION

**Corresponding Authors**

*E-mail: kerstin.goepfrich@mr.mpg.de

*E-mail: spatz@mr.mpg.de

*E-mail: na.liu@kip.uni-heidelberg.de

**Author Contributions**

The manuscript was written through contributions of all authors. All authors have given approval to the final version of the manuscript. -,rThese authors contributed equally.

**Notes**

The authors declare no competing financial interests.

# ABBREVIATIONS

AuNR, gold nanorod; CD, circular dicroism; TEM, transmission electron microscopy; RH, right-handed; LH, left-handed

# ACKNOWLEDGEMENT

K.G., C.F., I.P. and J.P.S. acknowledge funding from the European Research Council, Grant Agreement no. 294852, SynAd and the MaxSynBio Consortium, which is jointly funded by the Federal Ministry of Education and Research of Germany and the Max Planck Society. They also acknowledge the support from the SFB 1129 of the German Science Foundation and the VolkswagenStiftung (priority call Life). J.P.S. is the Weston Visiting Professor at the Weizmann Institute of Science and part of the excellence cluster CellNetworks at the University of Heidelberg. K.G. received funding from the European Union's Horizon 2020 research and innovation program under the Marie Sk-lodowska-Curie grant agreement No 792270, by the Deutsche Forschungsgemeinschaft (DFG, German Research Foundation) under Germany's Excellence Strategy via the Excellence Cluster 3D Matter Made to Order (EXC-2082/1 - 390761711) and the Max Planck Society. M.J.U and N.L were supported by the European Research Council (ERC Dynamic Nano). M.J.U. acknowledges the financial support by the Carl-Zeiss-Stiftung. We thank Marion Kelsch for assistance with TEM. TEM images were collected at the Stuttgart Center for Electron Microscopy. We thank X. Shen for synthesis of the AuNRs. The Max Planck Society is appreciated for its general support.

# TOC Graphic

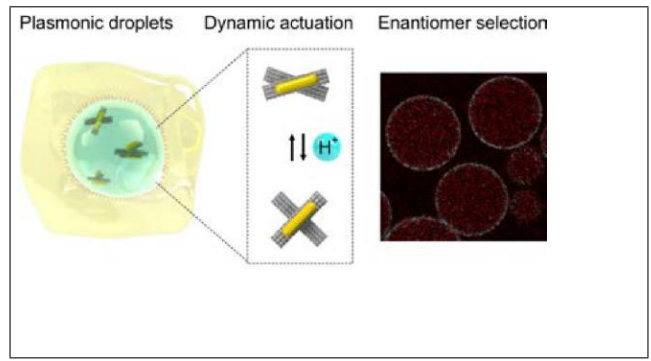